\documentclass{article}
\usepackage{spconf,amsmath,graphicx}
\usepackage{booktabs}
\usepackage{multirow}
\usepackage{threeparttable, url}
\usepackage{enumitem}
\usepackage{CJKutf8}

\title{The Database and Benchmark for the Source Speaker Tracing Challenge 2024}
\name{
\parbox{\linewidth}{\centering
Ze Li$^{1,2}$, Yuke Lin$^{1,2}$, Tian Yao$^{3}$, Hongbin Suo$^{3}$, Pengyuan Zhang$^{4}$ ,Yanzhen Ren$^{5}$, Zexin Cai$^{6}$,\\
Hiromitsu Nishizaki$^{7}$, Ming Li$^{1,2}$ 
}}

\address{
$^1$School of Computer Science, Wuhan University, Wuhan, China \\
$^2$Suzhou Municipal Key Laboratory of  Multimodal Intelligent Systems,\\ Duke Kunshan University, Kunshan, China \\
$^3$AI Center, OPPO, Beijing, China \\
$^{4}$ Key Laboratory of Speech Acoustics and Content Understanding,Institute of Acoustics,CAS, China \\
$^{5}$ Key Laboratory of Aerospace Information Security and Trusted Computing, Ministry of Education, \\
School of Cyber Science and Engineering, Wuhan University \\
$^{6}$ Center for Language and Speech Processing, Johns Hopkins University, USA \\
$^{7}$ Integrated Graduate School of Medicine, Engineering, and Agricultural Sciences, \\
University of Yamanashi, 4-4-37, Takeda, Kofu, 400-8510, Yamanashi, Japan\\
ming.li369@dukekunshan.edu.cn
}

\begin{document}
\begin{CJK}{UTF8}{gbsn}
\ninept
\maketitle
\begin{abstract}
    Voice conversion (VC) systems can transform audio to mimic another speaker's voice, thereby attacking speaker verification (SV) systems. However, ongoing studies on source speaker verification (SSV) are hindered by limited data availability and methodological constraints. This paper presents the Source Speaker Tracking Challenge (SSTC) on STL 2024, which aims to fill the gap in the database and benchmark for the SSV task. In this study, we generate a large-scale converted speech database with 16 common VC methods and train a batch of baseline systems based on the MFA-Conformer architecture. In addition, we introduced a related task called conversion method recognition, with the aim of assisting the SSV task. 
    We expect SSTC to be a platform for advancing the development of the SSV task and provide further insights into the performance and limitations of current SV systems against VC attacks. Further details about SSTC can be found here \footnote{https://sstc-challenge.github.io/}.
\end{abstract}
\section{Introduction}
\label{sec:intro}
Automatic Speaker Verification (ASV) systems aim to verify the identities of speakers from their voice samples. In recent years, ASV has been used in many applications, along with advancements in deep neural networks. Deep learning based ASV systems \cite{x-vector, ecapa} have demonstrated impressive performance across various scenarios. However, the development of synthetic speech technology poses a significant threat to the security of ASV systems. Advanced VC techniques can generate realistic fake voices, making deep learning based ASV systems vulnerable to spoofing attacks.

\begin{figure}
  \centering
  \includegraphics[width=\linewidth]{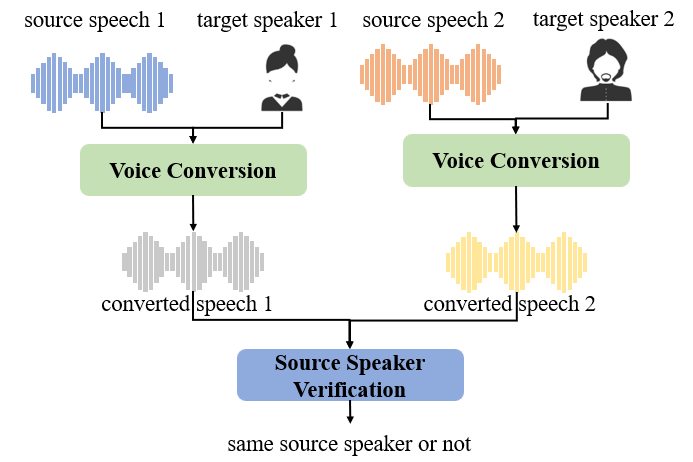}
  \caption{Source speaker verification against voice conversion. }
  \label{fig:ssv}
\end{figure}

VC technology involves transforming the original speaker's speech into speech with a new target speaker's timbre while maintaining the content unchanged. In recent years, scholars and research institutions have organized and conducted the Automatic Speaker Verification Spoofing and Countermeasures (ASVspoof) Challenges \cite{asvspoof2015,asvspoof2017,asvspoof2019,asvspoof2021} and Audio Deepfake Detection (ADD) Challenges \cite{add2022, add2023} series of competitions and developed various defense strategies \cite{spoofing_suvery, spoof_vc} to combat VC spoofing attacks. However, these strategies are typically used to distinguish between genuine and spoofed speech and do not consider the functionality of identifying the attacker (source speaker) in the converted speech. The SSTC aims to address this gap by focusing on identifying the information of the source speaker in manipulated speech signals. As depicted in Fig. \ref{fig:ssv}, while VC alter the source speech signal to mimic target speaker, the SSV system should extract information about the source speaker from the converted speech and decide whether two converted utterances are from the same source speaker.


\begin{table*}[htbp]\centering \scriptsize 
    \caption{Train, dev and test sets and repositories for each VC method.}
     \label{tab:database}
    \begin{threeparttable}
    \begin{tabular}{lccccccl}
    \toprule
    \multirow{2}*{\textbf{Method}} & \multicolumn{2}{c}{\textbf{Train set}} & \multicolumn{2}{c}{\textbf{Dev set}} & \multicolumn{2}{c}{\textbf{Test set}} & \multirow{2}*{\textbf{Repository}}\\
    \cmidrule(lr){2-3} \cmidrule(lr){4-5} \cmidrule(lr){6-7}& \textbf{ID} & \textbf{\#Utterances} & \textbf{ID} & \textbf{\#Utterances} & \textbf{ID} & \textbf{\#Utterances} \\
    \midrule 
    AGAIN-VC    & Train-1 & 327,600 & Dev-1 & 14,622 & Test-1 & 13,530 &KimythAnly/AGAIN-VC \\
    FreeVC      & Train-2 & 327,561 & Dev-2 & 14,622 & Test-2 & 13,530 &OlaWod/FreeVC \\
    MediumVC    & Train-3 & 327,609 & Dev-3 & 14,622 & Test-3 & 13,530 &BrightGu/MediumVC\\
    StyleTTS    & Train-4 & 327,546 & Dev-4 & 14,622 & Test-4 & 13,530 &yl4579/StyleTTS-VC\\
    TriAAN-VC   & Train-5 & 327,609 & Dev-5 & 14,622 & Test-5 & 13,530 &winddori2002/TriAAN-VC\\
    VQMIVC      & Train-6 & 327,498 & Dev-6 & 14,622 & Test-6 & 13,530 &Wendison/VQMIVC\\
    SigVC       & Train-7 & 327,603 & Dev-7 & 14,622 & Test-7 & 13,530 &- \\
    KNN-VC      & Train-8 & 327,765 & Dev-8 & 14,622 & Test-8 & 13,530 &bshall/knn-vc\\
    BNE-PPG-VC  & - & - & Dev-9   & 14,622 & Test-9 & 13,530 &liusongxiang/ppg-vc\\
    DiffVC      & - & - & Dev-10  & 14,622 & Test-10 & 13,530 &huawei-noah/Speech-Backbones\\
    S2VC        & - & - & Dev-11  & 14,622 & Test-11 & 13,530 &howard1337/S2VC\\
    YourTTS     & - & - & Dev-12  & 14,622 & Test-12 & 13,530 &Edresson/YourTTS\\
    ControlVC   & - & - & - & - & Test-13 & 13,530 & MelissaChen15/control-vc\\
    Diff-HierVC & - & - & - & - & Test-14 & 13,530 & hayeong0/Diff-HierVC\\
    LVC-VC      & - & - & - & - & Test-15 & 13,530 & wonjune-kang/lvc-vc\\
    Wav2vec-VC  & - & - & - & - & Test-16 & 13,530 & prairie-schooner/wav2vec-vc\\
    \bottomrule
    \end{tabular}
    \begin{tablenotes}
        \item * All the repositories can be retrieved in the \url{https://github.com}.
    \end{tablenotes}
    \end{threeparttable}
\end{table*}

Cai et al.'s research \cite{cdw} indicates that existing VC techniques are imperfect, and the converted speech retains some aspects of the source speaker's speech style. They astutely captured this observation and successfully introduced the identification of the source speaker by training converted audio with source speaker labels for SSV.

Since the concept of SSV is new, no open-source datasets are available to support this task. In this paper, we construct a large-scale database of converted speech by 16 well-performed VC methods, i.e., AGAIN-VC \cite{againvc}, FreeVC \cite{freevc}, MediumVC \cite{mediumvc}, StyleTTS \cite{styletts}, TriAAN-VC \cite{triaan}, VQMIVC \cite{vqmivc}, KNN-VC \cite{knnvc}, SigVC \cite{sigvc}, BNE-PPG-VC \cite{bne-ppg-vc}, DiffVC \cite{diffvc}, S2VC \cite{s2vc}, YourTTS \cite{yourtts}, ControlVC \cite{controlvc}, Diff-HierVC \cite{diffhiervc}, LVC-VC \cite{lvc} and Wav2vec-VC \cite{wav2vec-vc}. We train a batch of baseline systems using the MFA-Conformer \cite{mfa} model with the converted speech database. 

In addition to the source speaker information, the converted speech also contains traces of conversion methods. 
Due to essential differences among different VC methods, we can extract these conversion traces to identify the conversion method and to further process the converted speech. Based on this view, we introduce a related task—conversion method recognition \cite{add2023}. We employ a multi-task learning approach using an Adapter-based MFA-Conformer model \cite{half_mfa} for both SSV and conversion method recognition. Since obtaining training sets for all possible conversion methods is not feasible, we extend the closed-set conversion method recognition task to the open-set conversion method recognition task and employ an open-set nearest neighbor (OSNN) \cite{osnn} method for evaluation.

We summarize our main contributions as follows:
\begin{itemize}[leftmargin=*]
    \item We generate and release a large-scale converted speech database with 16 common VC methods for the SSV task. 
    \item We train a batch of baseline systems using the MFA-Conformer architecture on the large-scale converted speech database and \mbox{establish} benchmarks.
    \item We introduce a related task called conversion method recognition and employ multi-task learning with OSNN approach to \mbox{simultaneously} tackle the SSV and the open-set conversion method recognition tasks.
\end{itemize}

\section{Database}
\label{sec:db}

\begin{figure*}
  \centering
  \includegraphics[width=\linewidth]{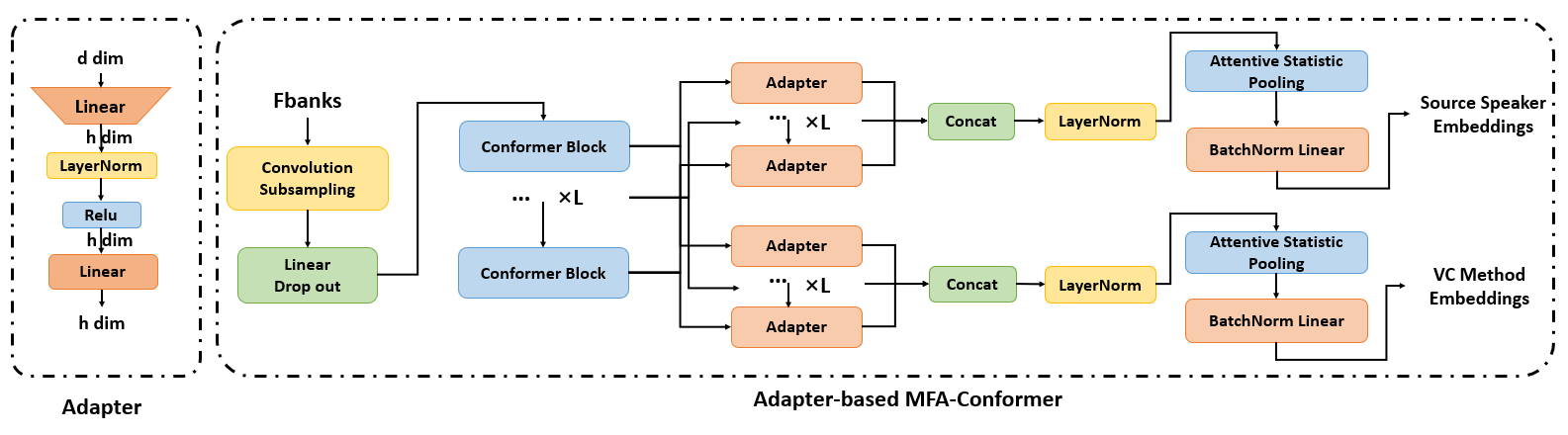}
  \caption{Adapter and Adapter-based MFA-conformer model structure.}
  \label{fig:system_framework}
\end{figure*}

As shown in Table \ref{tab:dataset_s_t}. We utilize Librispeech \cite{librispeech} as the source speaker dataset and VoxCeleb \cite{voxceleb, voxceleb2} as the target speaker dataset. Within Librispeech, the train-clean section, comprising 132,553 utterances from 1,172 speakers, is chosen as the training set. The dev-clean subset comprising 2,703 utterances from 40 speakers is designated as the development set. While the test-clean subset, with 2,620 utterances from 40 speakers, serves as the test set. In VoxCeleb, our training data is sourced from the VoxCeleb2 development set, encompassing 1,092,009 utterances from 5,994 speakers. For the development set, we utilize the VoxCeleb1 test set, consisting of 4,847 utterances from 40 speakers. A subset of the VoxCeleb1 development set, composed of 4,510 utterances from 40 speakers, serves as the test set.

\begin{table}[htbp]\centering \scriptsize
    \caption{Source speaker dataset and target speaker dataset.}
     \label{tab:dataset_s_t}
    \begin{tabular}{lccc}
    \toprule
    \textbf{Dataset} & \textbf{Subset} &\textbf{Speakers} & \textbf{\#Utterances} \\
    \midrule 
      & Train & 1,172 & 132,553 \\
     Source Speaker & Dev & 40 & 2,703 \\
      & Test & 40 & 2,602\\
     \midrule
      & Train & 5,994 & 1,092,009 \\
     Target Speaker & Dev & 40 & 4,847 \\
      & Test & 40 & 4510 \\
    \bottomrule
    \end{tabular}
\end{table}

We adopt the method proposed by Cai \cite{cdw} to generate converted speech. For each target speech, three source speech samples are randomly selected for voice conversion, simulating attacks from three distinct attackers on the same target speech. To optimize storage efficiency without compromising training data diversity, we partition the VoxCeleb training set into ten subsets of equal size while preserving the number of speakers. Each VC method utilizes one of these subsets as the target speech set for generating converted speech. The total number of converted speech samples for training in each VC method is approximately 327,600 ($109,200\times3$). For development and testing, the number of converted speech samples for each VC method is 14,622 ($4847\times3$) and 13,530 ($4510\times3$), respectively.

We introduce 16 any-to-any VC methods to generate the large-scale converted speech dataset, as shown in Table \ref{tab:database}. For SigVC, there is no publicly available repository, and we use the in-house implementation. The details of the SigVC model can be found in \cite{sigvc}. For other VC methods, we use the pre-trained VC models and the official implementation.

For evaluation, We divide the enrollment and test utterance into four scenarios: (1) the same source speaker and the same target speaker, (2) the different source speakers and the same target speaker, (3) the same source speaker and the different target speakers, and (4) the different source speakers and the different target speakers. We randomly generate enrollment and test pairs according to these four scenarios and ensure that the number of pairs for each scenario is the same to create a balanced set of trials. The total number of development and test sets trials is 350,928 and 324,720, respectively.

\section{Evaluation Protocol}
We use the Equal Error Rate (EER) metric to evaluate the system's performance in this challenge. For each pair of two converted speech utterances in the development and test sets, the cosine similarity is computed, and the decision on same-source-speaker vs. different-source-speakers is made by threshold. Denoting by $P_{fa}(\theta)$ and $P_{miss}(\theta)$ the false alarm and miss rates at threshold θ, the $EER$ metric corresponds to the threshold $\theta_{EER}$ at which the two detection error rates are equal, i.e., $EER$ = $P_{fa}(\theta_{EER})$ = $P_{miss}(\theta_{EER})$. We will compute the average $EER$ for all test sets as the final evaluation criteria:
\begin{equation}
    \textbf{$Score$} = \frac{1}{N} \sum_{i=1}^{N} \text{$EER_{i}$}
\end{equation}

\noindent where $N$ is the total number of test sets, which is 16 in this challenge. $EER_{i}$ represents the $EER$ value of the $i$-th test set. The lower the $Score$, the greater the system performance.

\section{Baseline System}
\label{sec:baseline}
This section provides a detailed description of our MFA-Conformer-based baseline system. Additionally, we explore a related task: conversion method recognition. Incorporating Adapter modules into the baseline system enables effective conversion method recognition without compromising the SSV performance. Furthermore, we introduce an OSNN approach to effectively address the open-set conversion method recognition problem. The code can be found here\footnote{https://github.com/SSTC-Challenge/SSTC2024\_baseline\_system}.

\subsection{Source Speaker Embedding Clustering}
Speaker embedding networks aim to learn a discriminative embedding space that clusters utterances from the same speaker together while separating those from different speakers. VC models enable a source voice to sound like that of a target speaker, altering the embedding distribution of the source voice to fall within the target speaker's subspace, thereby deceiving speaker verification systems. The goal of source speaker identification is to learn a speaker embedding space that maps the converted voice back to the source speaker's subspace.

Given a source speech dataset $D_{s}$ and a target speech dataset $D_{t}$, the VC model $V$ operates on the source speech $s_{i} \in D_{s}$ to make it sound like the target speech $t_{j} \in D_{t}$, thus generating a converted speech dataset $D_{v} = \{ V_{s_{i} \rightarrow t_{j}} \mid s_{i} \in D_{s}, t_{j} \in D_{t} \}$. All converted speech datasets $D_{v,k}$ associated with VC methods $V_{k} (k=1,2,..., K)$ are combined as training data, and the label for the converted speech $V_{s_{i} \rightarrow t_{j}}$ is set to the speaker identify of the source speech $s_{i}$ to enable the speaker embedding network to learn the information of the source speaker from the converted speech. It allows the network to map the converted speech back to the subspace of the source speaker while maintaining a discriminative speaker embedding space.

\subsection{Adapter-based Multi-Task Learning}
While the objective of SSV and conversion method recognition differs, these tasks are interrelated since the converted speech sample contains both source speaker information and information about the speech conversion method. To achieve dual objectives in a single stroke, we design a multi-task learning approach to concurrently identify the source speaker and the specific method.

In multi-task learning, adapters \cite{adapter1, adapter2} are a common technique to simultaneously learn multiple related but distinct tasks within a single model. The fundamental idea of adapters is to introduce tiny, task-specific parameter sets between different layers of the model, allowing parameter sharing across different tasks and facilitating a multi-task learning process with minimum cost.

As shown in Fig \ref{fig:system_framework}, we employ the Adapter-based MFA-Conformer model\cite{half_mfa} by adding an adapter after each conformer block in the MFA-Conformer architecture. 

These adapters fine-tune the outputs from each layer of the conformer model and align them more closely with the target task.
In detail, the frame-level output from the $i^{th}$ conformer layer, denoted as $ h_{i} \in R^{d \times T}$, is transformed by the adaptor $A_{i}$, and then we concatenate the output feature maps from each adapter and feed them into a LayerNorm layer:
\begin{equation}
    h_{i}^{'} = A_{i}(h_{i}) , \quad i = 1, 2, \ldots, L
\end{equation}
\begin{equation}
    H = Concat(h_{1}^{'},h_{2}^{'},...,h_{L}^{'})
\end{equation}

\subsection{Open-set voice conversion methods recognition}
While recognizing VC methods present in the training set is straightforward, the real challenge lies in distinguishing the unseen methods. This is due to the impossibility of encompassing all VC techniques globally within our training data, making it a significant hurdle to overcome.

\begin{figure}
  \centering
  \includegraphics[width=\linewidth]{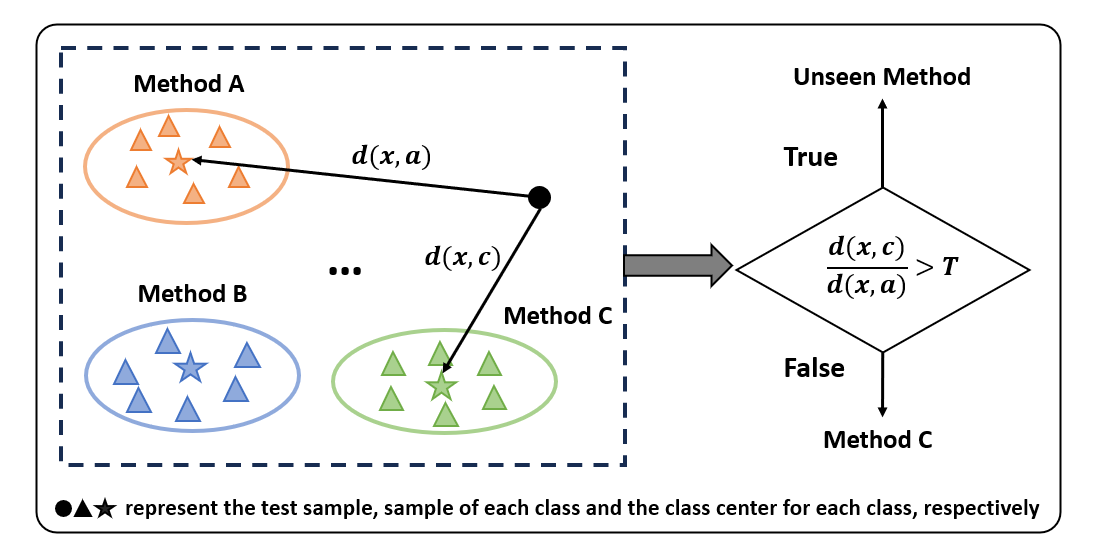}
  \caption{Open-set VC methods recognition based on OSNN method.}
  \label{fig:osnn}
\end{figure}

During training, utterances from the same VC method tend to cluster together, while those from different methods are scattered. Therefore, audio samples of VC methods in the training set will be similar to samples from known categories. In contrast, audio samples from unseen methods will be different from samples from known categories. Furthermore, similarity can be calculated by comparing the Euclidean distances to different categories. Based on this theory, we introduce the OSNN classification method, as shown in Fig \ref{fig:osnn}. The steps of this method are as follows:

\begin{itemize}[leftmargin=*]
    \item Step 1. Extract all method embeddings $\mathbf{Z} \in R^{N \times d}$ from the training set with the trained method recognition model and randomly partition them into two subsets at a ratio of 1:9, denoted as $TS_{1}$ and $TS_{9}$.
    \item Step 2. Take the subset $TS_{9}$ and calculate the average of embeddings for each method to obtain the class center of each method.
    \item Step 3. Calculate the Euclidean distance from test sample $\mathbf{x}$ to each class center, and compute the ratio of distances $R$ between its two nearest neighbor centers $\bf c_i$ and $\bf c_j$. In which, $\bf c_i$ represents the nearest neighbor, and $\bf c_j$ represents the second nearest neighbor: 
        \begin{equation}
            d(\mathbf{x}, \mathbf{y}) = \sqrt{\sum_{i=1}^{n} (\mathbf{x}_i - \mathbf{y}_i)^2}
        \end{equation}
        \begin{equation}
            R = d(\mathbf{x}, \mathbf{c_i})/d(\mathbf{x},\mathbf{c_j})
        \end{equation}
    \item Step 4: Given a specified threshold $T$, if the distance ratio $R$ is less than the threshold, consider the test sample belonging to method $i$; otherwise, consider the sample belonging to an unseen method.
\end{itemize}

\begin{figure}
  \centering
  \includegraphics[width=0.95\linewidth]{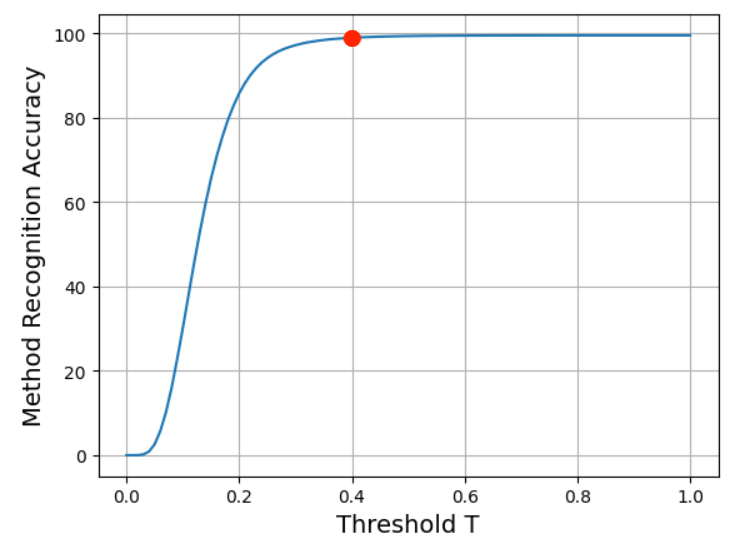}
  \caption{The accuracy of conversion method recognition on the subset $TS_{1}$ using different thresholds $T$ (0 \textless $T$ \textless 1).}
  \label{fig:t}
\end{figure}

The threshold $T$ is determined by the subset $TS_{1}$. We gradually increase the value of $T$ from 0 to 1 and perform Step 3 and Step 4 on this part of the data. We then calculate the average recognition accuracy at different thresholds. Finally, we select the threshold where the accuracy begins to stabilize as the final threshold $T$. As shown in Fig \ref{fig:t}, the recognition accuracy increases rapidly as the threshold increases from 0 to 0.4. However, after the threshold reaches 0.4, the increase in accuracy becomes less pronounced. Therefore, we choose 0.4 as the final value for the threshold $T$.

\subsection{Experimental Setup}

\begin{table*}[htbp]\centering \scriptsize
    \caption{The EER results of SSV on test sets based on the half small MFA-Conformer system with different training sets and different training strategies. VC-2 represent training set mixed by two training sets of Train-1 and Train-2. VC-4 represent training set mixed by four training sets of Train-1, Train-2, Train-3 and Train-4. VC-8 represent training set mixed with all 8 training sets of converted speech.}
    \tabcolsep=0.25em
     \label{tab:result}
\begin{threeparttable}
    \begin{tabular}{lcccccccccccccccc}
    \toprule

    ~ &\textbf{Test-1} &\textbf{Test-2}&\textbf{Test-3}&\textbf{Test-4}&\textbf{Test-5}&\textbf{Test-6}&\textbf{Test-7}&\textbf{Test-8} &\textbf{Test-9}&\textbf{Test-10}&\textbf{Test-11}&\textbf{Test-12}&\textbf{Test-13}&\textbf{Test-14}&\textbf{Test-15}&\textbf{Test-16}
    \\
    \midrule 
     Train-1 & \textbf{8.960\%} & 45.931\% & 35.740\% & 43.706\% & 45.188\% & 43.968\% & 49.573\% & 50.462\% & 49.611\% & 50.506\% & 39.113\% & 48.745\% & 47.029\% & 44.780\% & 47.440\% & 42.655\%    \\
     Train-2 & 38.158\% & \textbf{9.618\%} & 28.203\% & 33.837\% & 28.865\% & 36.025\% & 45.152\% & 43.851\% & 42.293\% & 46.673\% & 28.450\% & 28.347\% & 39.098\% & 17.660\% & 11.699\% & 39.144\% \\
     Train-3 & 32.581\% & 31.099\% & \textbf{6.901\%} & 33.927\% & 29.712\% & 35.099\% & 45.755\% & 44.842\% & 45.208\% & 47.539\% & 20.751\% & 32.409\% & 43.354\% & 27.498\% & 15.511\% & 35.201\% \\
     Train-4 & 34.526\% & 37.417\% & 32.877\% & \textbf{6.609\%} & 33.750\% & 39.145\% & 46.274\% & 46.734\% & 45.714\% & 48.128\% & 34.941\% & 38.169\% & 45.592\% & 32.579\% & 23.257\% & 39.512\%  \\
     Train-5 & 27.388\% & 25.410\% & 20.774\% & 28.919\% & \textbf{5.599\%} & 31.725\% & 45.691\% & 44.204\% & 42.340\% & 46.905\% & 19.144\% & 26.008\% & 42.832\% & 18.274\% & 9.438\% & 32.634\%  \\
     Train-6 & 40.260\% & 30.305\% & 29.979\% & 35.634\% & 33.399\% & \textbf{11.452\%} & 46.111\% & 44.909\% & 39.039\% & 46.401\% & 30.780\% & 34.450\% & 39.507\% & 29.650\% & 23.333\% & 41.412\% \\
     Train-7 & 42.368\% & 33.070\% & 39.270\% & 40.841\% & 37.552\% & 37.172\% & \textbf{30.472\%} & 43.056\% & 39.634\% & 43.025\% & 36.527\% & 36.415\% & 39.550\% & 34.330\% & 29.881\% & 40.110\% \\
     Train-8 & 35.613\% & 22.854\% & 29.546\% & 33.096\% & 26.800\% & 35.148\% & 40.196\% & \textbf{29.277\%} & 36.350\% & 40.745\% & 27.080\% & 26.783\% & 34.466\% & 23.538\% & 18.931\% & 35.471\% \\
     VC-2 & \textbf{9.222\%} & \textbf{9.687\%} & 25.583\% & 32.972\% & 27.055\% & 36.385\% & 44.805\% & 43.513\% & 42.107\% & 46.396\% & 25.816\% & 27.765\% & 38.718\% & 17.008\% & 11.455\% & 35.142\% \\
     VC-4 & \textbf{9.480\%} & \textbf{10.403\%} & \textbf{6.967\%} & \textbf{7.131\%} & 25.469\% & 36.176\% & 44.136\% & 43.032\% & 43.272\% & 46.782\% & 24.220\% & 27.685\% & 38.529\% & 24.015\% & 12.015\% & 32.867\% \\
     VC-8 & \textbf{9.786\%} & \textbf{10.645\%} & \textbf{6.999\%} & \textbf{7.606\%} & \textbf{6.732\%} & \textbf{10.756\%} & \textbf{32.902\%} & \textbf{29.303\%} & 34.593\% & 45.415\% & 18.714\% & 22.501\% & 36.657\% & 20.368\% & 9.530\% & 27.308\% \\
     \midrule 
     Multi-VC-8 & \textbf{14.621\%} & \textbf{15.465\%} & \textbf{10.427\%} & \textbf{11.679\%} & \textbf{11.602\%} & \textbf{16.321\%} & \textbf{38.805\%} & \textbf{37.402\%} & 47.166\% & 46.375\% & 21.113\% & 45.182\% & 37.430\% & 32.021\% & 12.464\% & 29.072\%   \\
     + Adapter & \textbf{9.950\%} & \textbf{10.402\%} & \textbf{6.943\%} & \textbf{7.636\%} & \textbf{6.683\%} & \textbf{10.783\%} & \textbf{32.392\%} & \textbf{28.418\%} & 33.638\% & 44.209\% & 19.396\% & 22.393\% & 36.272\% & 19.902\% & 9.561\% & 27.258\%   \\
    \bottomrule
    \end{tabular}
\end{threeparttable}
\end{table*}

\subsubsection{Model usage}
In addition to employing the MFA-Conformer \cite{mfa} model for training, we also utilized the ResNet34 \cite{resnet} architecture for comparative analysis. For ResNet34, we configure the channels of residual blocks as \{64,128,256,512\}. The ResNet’s output feature maps are aggregated with a global statistics pooling layer that calculates each feature map's means and standard deviations. For the MFA-Conformer, we reduced the number of Conformer layers by half while increasing the sampling rate from 1/4 to 1/2 \cite{half_mfa}, building upon using a small-sized conformer encoder. 
The adapter structure consists of two linear layers followed by Layer Normalization and Rectified Linear Unit (ReLU) activation function as shown in Fig \ref{fig:system_framework}. The first linear layer projects the input data from dimension $d$ to a hidden layer size of 128. Another linear layer maps the 128-dimensional feature representation to another 128-dimensional space. The acoustic features are 80-dimensional log Mel-filterbank energies with a frame length of 25ms and a hop size of 10ms. The extracted features are mean-normalized before feeding into the deep speaker network.

\subsubsection{Training details}
We adopt the on-the-fly data augmentation \cite{on-the-fly} to add additive background noise or convolutional reverberation noise for the time-domain waveform. The MUSAN \cite{musan} and RIR Noise \cite{RIR} datasets are used as noise sources and room impulse response functions, respectively. To further diversify training samples, we apply amplification or playback speed change (pitch remains untouched) to audio signals. 

Network parameters are updated using AdamW optimizer \cite{adamw} with cosine decay learning rate schedule. The initial learning rate starts from 1.0e-3, and the minimum learning rate is 1.0e-5. We perform a linear warm-up learning rate schedule at the first epoch to prevent model vibration and speed model training. The input frame length is fixed at 200 frames. For the SSV task, we employ the ArcFace classifier, with the margin and scale parameters set as 0.2 and 32, respectively. For the conversion method recognition task, we utilize a linear classifier. Additionally, we employ a pre-training strategy on the VoxCeleb2 development set to allow the model to learn general features and perform better on subsequent tasks.

\section{Results and analysis}
\label{sec:res}

\subsection{Baseline System Results and Analysis}

Table \ref{tab:ResNet-MFAConformer} presents a performance comparison of different model architectures on the SSV task. In this experiment, both models are trained exclusively on the Train-1 training set and tested on the Dev-1 and Test-1 sets. It can be observed that, despite the MFA-Conformer model has only 8.68M parameters, which is two-fifths of ResNet34, its EER on the Dev-1 and Test-1 sets reaches 9.412\% and 9.424\%, respectively, significantly better than the ResNet34's 13.01\% and 13.463\%. 
This improvement can be attributed to Conformer's attention-based sequence modeling architecture, which facilitates better focus on capturing source speaker information in converted speech while disregarding interference from target speaker information.
Additionally, the strategy of initializing model parameters with pre-trained weights from VoxCeleb2 further enhances performance, with the MFA-Conformer model achieving EERs of 8.48\% on Dev-1 set and 8.96\% on Test-1 set.

\begin{table}[htbp]\centering \scriptsize
    \caption{Comparison of performance between ResNet34-GSP and MFA-Conformer half small. Trained solely with training set Train-1, tested on the Dev-1 and Test-1 .}
     \label{tab:ResNet-MFAConformer}
    \begin{tabular}{lccc}
    \toprule
    \textbf{Model} & \textbf{Parameters} &\textbf{Dev-1} &\textbf{Test-1}\\
    \midrule 
     ResNet34-GSP &  21.54M & 13.010\% & 13.463\% \\
     + VoxCeleb2 pre-train &  & 10.680\% & 10.673\% \\
     \midrule
     MFA-Conformer half small & 8.68M & 9.412\% & 9.424\%  \\
     + VoxCeleb2 pre-train &  &  8.480\% & 8.960\% \\
    \bottomrule
    \end{tabular}
\end{table}

Tabel \ref{tab:result} reports the SSV results on test sets based on the half small MFA-Conformer system. The performances are displayed from two aspects: different training sets and strategies. 

As we can see, when training and testing with speech generated from the same voice conversion (VC) model, the systems trained on Train-1$\sim$Train-6 achieved favorable EER results, specifically 8.960\%, 9.618\%, 6.901\%, 6.609\%, 5.599\%, and 11.452\%, respectively. However, the systems trained on Train-7 and Train-8 yielded significantly higher EER results of 30.472\% and 29.277\%, respectively. 
For SigVC, a speaker information removal module (SI Remover) is specifically designed to eliminate source speaker information to obtain purified speech content information.
For KNN-VC, it first extracts self-supervised features from both the source and target speech using WavLM\cite{wavlm}. Then, each source audio feature is replaced with the mean of the K closest target speech features. This means that the final converted speech features are entirely derived from the target audio features, with no features from the source speech, thereby eliminating the source speaker information.


Due to the inherent differences among various conversion methods, speech generated by different conversion methods can be considered data from different domains. When training data from different domains are mixed, it inevitably leads to a slight decrease in performance on each domain. We can observe that as the number of conversion methods increases, the EER of systems trained on Train-1, VC-2, VC-4, and VC-8 on the Test-1 set reduce, with values of 8.960\%, 9.222\%, 9.480\%, and 9.786\%, respectively. However, the system's generalization ability tends to improve when exposed to a greater variety of conversion types. For example, we can observe that systems trained on Train-1, VC-2, and VC-4 exhibit decreasing EER values on the Test-5 set, with values of 45.188\%, 27.055\%, and 25.469\%, respectively. Similar trends are evident across other test sets, suggesting an enhanced performance for previously unseen conversion methods. 
Nevertheless, there are some test sets where the EER consistently remains high, such as Test-10. This can be attributed to the conversion method DiffVC used in Test-10, which employs an average voice encoder to decouple content and speaker information, and it utilizes a Diffusion Probabilistic Model as its decoder, marking a fundamental difference from the models used in Test-1$\sim$Test-8.

\begin{figure*}
  \centering
  \includegraphics[width=0.9\linewidth]{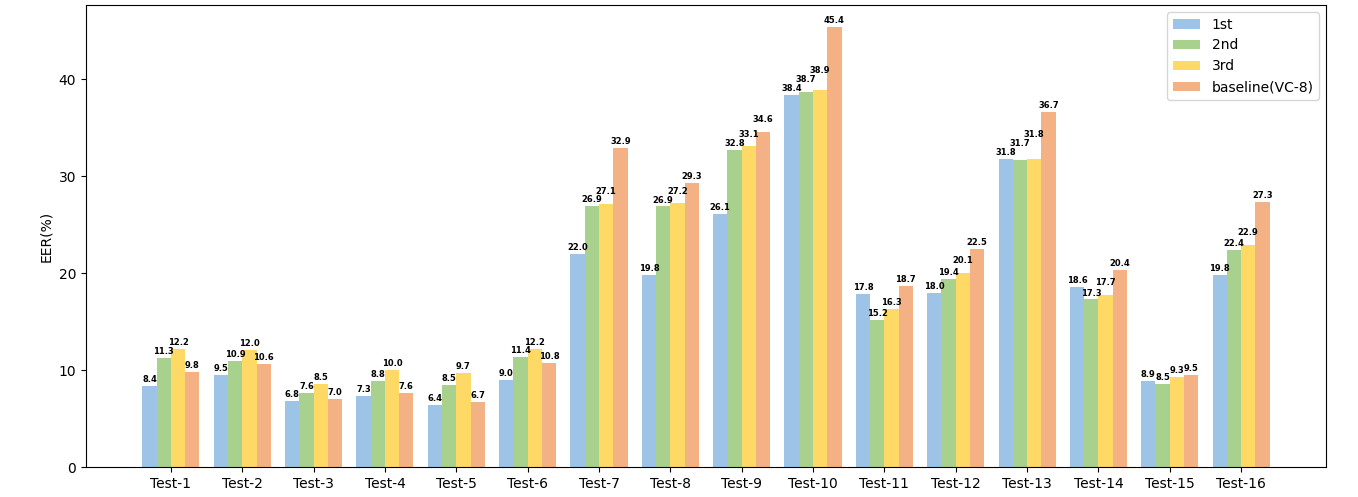}
  \caption{The EER results of SSTC2024's 1st place, 2nd place, 3rd place and baseline system VC-8 on different test sets.}
  \label{fig:res}
\end{figure*}

\begin{table}[htbp]\centering \scriptsize
    \caption{Average method recognition accuracy of the seen and unseen VC methods using the OSNN approach on the Multi-VC-8 with Adapter system.}
     \label{tab:method}
    \begin{tabular}{lcc}
    \toprule
    \textbf{Dataset} & \textbf{Average Accuracy on} &\textbf{Average Accuracy on} \\
    & \textbf{Seen Methods} & \textbf{Unseen Methods} \\
    \midrule 
     Dev sets&  98.36\% &  99.16\%  \\
     Test sets & 98.11\% & 90.68\% \\
    \bottomrule
    \end{tabular}
\end{table}


For multi-task learning, the differences among different tasks are apparent. 
System Multi-VC-8 extends the system VC-8 by simultaneously performing source speaker verification and method recognition on the extracted embeddings, its performance in source speaker verification significantly degrades. On the Test sets of each VC method, the EER experiences a performance decrease of 3-4 points. After we add the adapter module, due to the ability of the adapter to fine-tune the outputs of each layer of the consistency model and align them more closely with the target task. 
It's worth noting that the system, despite the additional task of method recognition, manages to maintain a performance level comparable to system VC-8 on the SSV task. This is a testament to the system's robustness and its ability to handle multiple tasks effectively. Furthermore, for the open-set method recognition problem, we employed the OSNN method. As shown in Table \ref{tab:method}, the system ultimately achieves an average recognition accuracy on Dev sets and Test sets of 98.36\% and 98.11\% on the seen methods, and 99.16\% and 90.68\% on the unseen methods.


\subsection{Technical Systems Results and Analysis}
SSTC2024 attracted a total of registration 50 teams. Of these, 8 teams submitted their results on the challenge leaderboard platform\footnote{https://codalab.lisn.upsaclay.fr/competitions/18512} during the evaluation phase. Among them, 6 teams provided final system descriptions to explain and verify the correctness of their systems. Teams, which do not submit system descriptions are excluded from the final rankings. The final rankings are shown in Table \ref{tab:codalab}. We can observe that the baseline system Multi-VC-8 with Adapter and VC-8 ranks fifth and sixth, the first-place team achieved a score of 16.788\%, with a 3.825\% improvement over the baseline VC-8. However, most participants' scores were not significantly better than the baseline, with some even achieving as poor as 54.422\%. This indicates that the SSV task still presents considerable challenges.

\begin{table}[htbp]\centering \scriptsize
    \caption{Final Rankings of SSTC2024. The Baseline System Ranked fifth and sixth.}
     \label{tab:codalab}
    \begin{tabular}{lcc}
    \toprule
    \textbf{Rank} & \textbf{User} & \textbf{Score} \\
    \midrule 
     1 & Rachel\_W & 16.788\% \\
     2 & zhangdejun & 18.648\% \\
     3 & wangzhi & 19.323\% \\
     4 & Masinless & 20.027\% \\
     \textbf{5} & \textbf{SSTC\_Organizer(Multi-VC-8 with Adapter)} & \textbf{20.365\%} \\
     \textbf{6} & \textbf{SSTC\_Organizer(VC-8)} & \textbf{20.613\%} \\
     
     7 & dkorzh10 & 20.669\% \\
     8 & akshetpatial & 54.422\% \\
    \bottomrule
    \end{tabular}
\end{table}

Fig. \ref{fig:res} provides a detailed overview of the scores achieved by the top three teams and the baseline system VC-8 across various test sets in the competition. The first-place team employed a source speaker contrastive learning method, effectively capturing the source speaker's information in the converted speech. They utilized the ResNet293 model architecture. The second-place team utilized a ResNet101 architecture along with an Adaptive Score Normalization (AS-Norm) \cite{as-norm} to enhance their scores. The third-place team employed a ResNet152 architecture and utilized embedding normalization to mitigate certain biases or noise in the data, thereby improving performance.
From Fig. \ref{fig:res}, it is evident that the first-place team consistently achieved the lowest EER across nearly all test sets, even achieving around 20\% EER on Test-7 and Test-8, representing approximately a 30\% performance improvement over the baseline. The second and third-place teams employed score calibration strategies, showing improvements over the baseline on Test-7$\sim$Test-16 but weaker performance on Test-1$\sim$Test-6. This aligns with previous findings that Conformer-based models excel in capturing source speaker information in converted speech with their attention-based sequence modeling architecture. In contrast, ResNet-based models benefit from their depth, allowing them to generalize better and learn more from the data.

\section{CONCLUSION}
This paper summarize the SSTC on SLT 2024. We generate and release a vast database of converted speeches with 16 common any-to-any VC methods for SSV task. Additionally, we train a set of baseline systems on this converted speech database using the MFA-Conformer-based architecture and establish benchmarks. Furthermore, we introduce a related task called conversion method recognition. Employing a multi-task learning approach with adapter-based method combined with OSNN, we successfully addressed both SSV and open-set conversion method recognition tasks simultaneously. We hope that future research can build on our work by utilizing accurately identified conversion methods to enhance the performance of SSV. 
The innovative methods and results of participating teams showcase the feasibility of contrastive learning and score calibration in the SSV task, while also highlighting that the SSV task remains exceedingly challenging with current technologies and has much room for improvement.

\section{ACKNOWLEDGMENTS}
\label{sec:ack}
This research is funded in part by the National Natural Science Foundation of China (62171207) and OPPO. Many thanks for the computational resource provided by the Advanced Computing East China Sub-Center.

\bibliographystyle{IEEEbib}
\bibliography{strings,refs}
\end{CJK}
\end{document}